\documentclass{PoS}

\title{Cascade Baryon Spectrum from Lattice QCD}

\ShortTitle{Cascade Baryons from Lattice QCD }
\author{ {\speaker{Nilmani Mathur}}$^a$, John M. Bulava$^b$, Robert G. Edwards$^c$, Eric Engelson$^d$, B\'alint Jo\'o$^c$, Adam Lichtl$^e$, Huey-Wen Lin$^c$, Colin Morningstar$^b$,  David G. Richards$^c$, and
         Stephen J. Wallace$^d$}
     \author{(for the Hadron Spectrum Collaboration)\\
\llap{$^a$}Department of Theoretical Physics, Tata Institute of Fundamental Research, Homi Bhabha Road, Mumbai 400005, India\\
        \llap{$^b$}Department of Physics, Carnegie Mellon University, Pittsburgh
, PA 15213\\
        \llap{$^c$}Thomas Jefferson National Accelerator Facility, Newport News,
 VA 23606\\
        \llap{$^d$}Department of Physics, University of Maryland, College Park, 
MD 20742\\
\llap{$^e$}RIKEN-BNL Research Center, Brookhaven National Laboratory, Upton, NY 11973\\
        E-mail: \email{nilmani@theory.tifr.res.in}}

\abstract{
 A comprehensive study of the cascade baryon spectrum using lattice
QCD affords the prospect of predicting the masses of states not yet
discovered experimentally, and determining the spin and parity of
those states for which the quantum numbers are not yet known. The
study of the cascades, containing two strange quarks, is particularly
attractive for lattice QCD in that the chiral effects are reduced
compared to states composed only of u/d quarks, and the states are
typically narrow. We report preliminary results for the cascade spectrum
obtained by using anisotropic $N_f$ = 2 Wilson lattices 
with temporal lattice spacing 5.56 GeV$^{-1}$.}

\FullConference{The XXVI International Symposium on Lattice Field Theory\\
		 July 14-19 2008\\
		 Williamsburg, Virginia, USA}

\begin{document}
\vspace*{-0.2in}
\section{Introduction}
A comprehensive understanding of the experimental hadron spectrum from
 a first principle non-perturbative calculation is one of the main
research program for lattice QCD. The lattice hadron spectrum
collaboration is mainly focusing on that. Along with nucleon spectrum
we are also extracting spectrum for hyperons, and cascade
baryons in particular. In this report we present preliminary findings for the
cascade baryons by using $N_f=2$ anisotropic Wilson action.

Cascade spectrum can be studied comprehensively by lattice QCD and
even with quenched lattice QCD. There are a few advantage of studying
cascade over studying nucleon, lambda, sigma and delta. First, at low
quark mass region, quenched QCD is plagued by the appearance of
unphysical ghost states arising out of $\eta^{\prime}$ loop. Since
these ghost states are associated with negative correlators, in
general, it is very difficult to extract any signal for a physical
state. However, since cascades have two strange-mass quarks, presences of
such unphysical states are somehow attenuated, although still present 
and most of the physical states appear as a ground states up to a quite
small pion mass region. With dynamical fermions there are no ghost states,
 but one has to
separate out multi-particle decay states from the resonance states. 
Secondly, due to the presence of two heavier strange quarks, the lattice signal for cascade baryons
 is much better than that of light quark baryons, such as the nucleon 
and delta.  Thirdly, again due to the presence of two strange quarks,
chiral extrapolation for cascade baryons are easier than many other
light quark baryons. Beside these, since the decay widths of cascades 
 are small there will be less ambiguity regarding the mixing of states.

\vspace*{-0.1in}

\section{Experimental cascade spectrum and lattice opportunity}
\begin{figure}
\begin{center}
\includegraphics[width=0.497\textwidth,height=0.35\textwidth,clip=true]{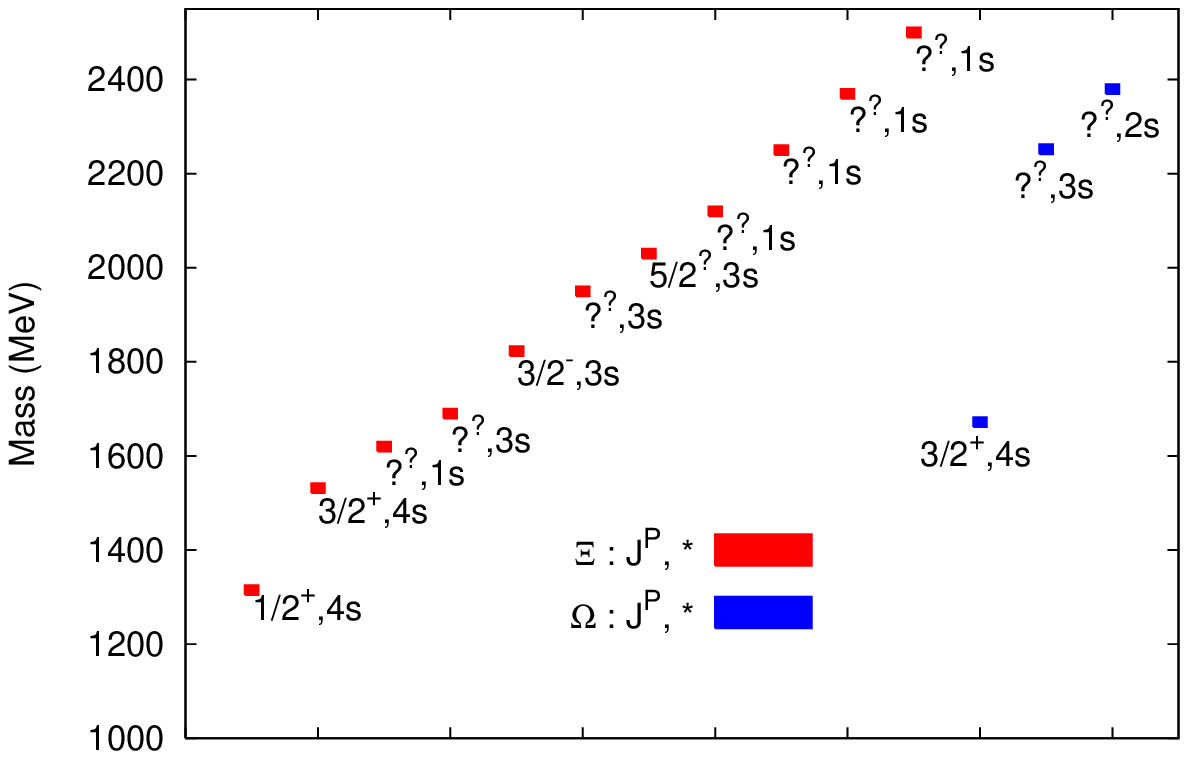}
\includegraphics[width=0.497\textwidth,height=0.35\textwidth,clip=true]{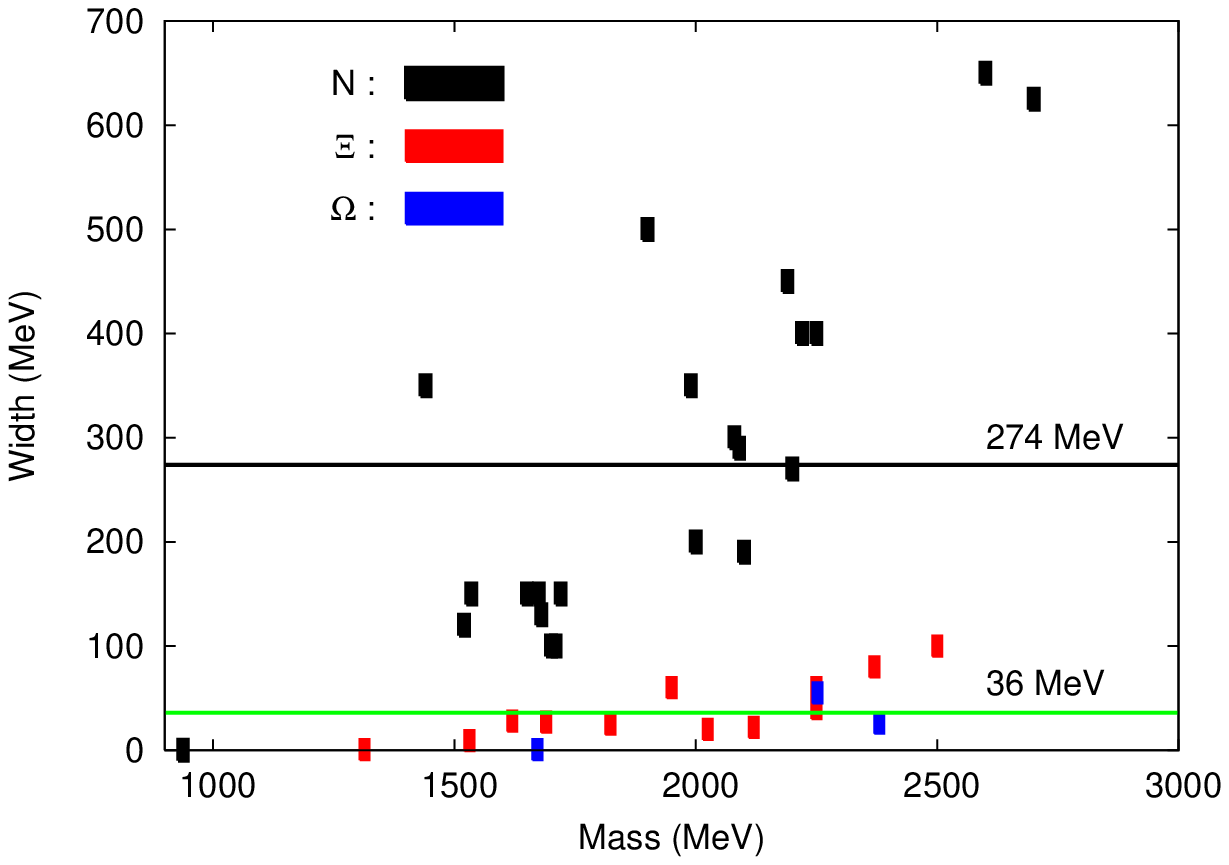} 
\end{center}
\vspace{-0.15in}
\caption{PDG quoted mass (left) and mean Breit-Wigner width (right) of $\Xi$ and $\Omega$ baryons. On the right figure mean Breit-Wigner widths of nucleon resonances are also plotted as a comparison. Combine average widths of $\Xi$ and $\Omega$'s, and average widths of nucleons are shown by green and black lines respectively.} \label{fig:mass_width}
\end{figure}

In Fig. \ref{fig:mass_width} we have plotted masses (left) and mean Breit-Wigner
widths (right) of experimental  cascade ($\Xi$) and omega ($\Omega$)
baryons. In the left figure known $\Xi$ and $\Omega$ states are shown along with
their quantum numbers ($J^{P}$) and star statuses (s). Except for a few
states (ground states in $1/2^{+}, 3/2^{+}, 3/2^{-}\,\, {\hbox{for}}
\,\, \Xi$ , and $3/2^{+}\, {\hbox{for}} \,\, \Omega $) either or both
of the quantum numbers of these states are not known. Another
interesting aspect of these baryons is that their decay widths are 
much narrower compare to nucleon states which are shown in right figure. The
average width of nucleon states is about 274 MeV, whereas for $\Xi$ and $\Omega$ baryons together it is just about 34 MeV.

By observing the experimental spectrum, as shown above, it is evident
that a comprehensive study of cascades and omegas by lattice QCD will
enable us to predict the quantum numbers of many states for which these are still unknown
experimentally. Also lattice predicted new states can be helpful for experimental discovery of those states. For example, one can hope to extract a reliable ground
state mass for negative parity octet cascade which is still not known.
 Also, it may be possible to extract a few excited states
reliably. For example, it is possible to extract the first excited
state in the positive parity octet channel. An indication of such a
state will open up the possibility of a Roper like state. 
That will settle the issue of the quantum numbers of the $\Xi(1690)$, where experimental evidence suggests a state of spin-1/2 \cite{Ziegler:2007tp}, but where the parity is still unresolved. Furthermore, a lattice prediction of two
states (negative parity ground state and positive parity excited
state) will be quite encouraging to search for these states
experimentally. The lattice prediction of excited states for spin 3/2
baryons will also settle quantum numbers for a few states around 1900 MeV, and can motivate the experimental search for the properties of such states.
Similarly for $\Omega$ baryons, one hopes to predict the negative parity
ground state and a positive parity excited state.  With our new sophisticated technology comprising group theoretical operators, as discussed below, we
 hope to predict many other excited states.

\vspace*{-0.1in}

\section{Cascade Operators}
As outlined in Ref. \cite{basak} for nucleons, we construct
cascade operators which transform irreducibly under the symmetries of the 
lattice i.e., of the octahedral group. 
In table 1 we show the irreducible representation (irreps) of
the lattice octahedral group $\Lambda=\{G_1,H,G_2\}$ and the
associated continuum angular momentum.
\begin{table}[h]
\caption{The irreducible representation of the octahedral group $\Lambda=\{G_1,H,G_2\}$ 
and the associated continuum angular momentum $J$ up to $\frac{13}{2}$.}
\begin{center}
\begin{tabular}{|c| c |}
\hline 
$\Lambda$ & $J$ \\
\hline 
G1  & $\frac{1}{2} \oplus  \frac{7}{2} \oplus  \frac{9}{2}  \oplus \frac{11}{2} \cdots$ \\
G2  & $\frac{5}{2} \oplus  \frac{7}{2} \oplus  \frac{11}{2} \oplus \frac{13}{2} \cdots$\\
H   & $\frac{3}{2} \oplus  \frac{5}{2} \oplus  \frac{7}{2}  \oplus \frac{9}{2}  \cdots$\\
\hline
 \end{tabular}
 \end{center}
 \label{tab:irreps}
 \end{table}

In order to build these operators one first constructs the basic
building blocks which are covariantly displaced smeared quark fields
$(\tilde D^{(n)}_j\tilde \psi(x))_{Aa\alpha}$ with flavour $A$, colour
$a$, and spin $\alpha$.  The smeared quark field $\tilde \psi$ is
displaced $n$ links in the $j$ direction ($j=0,\pm1,\pm2,\pm3$). It
has been observed that both quark smearing and gauge link smearing are
necessary to reduce noise and coupling to high energy states. As in
our previous studies \cite{adam}, we use Gaussian quark smearing and stout link
smearing. The next step is to construct elemental operators
from these displaced single quark operators as
\begin{equation}
\Phi^{ABC}_{\alpha\beta\gamma;ijk}(x) = \epsilon_{abc}(\tilde
D^{(n)}_i\tilde \psi(x))_{Aa\alpha}(\tilde D^{(n)}_j\tilde
\psi(x))_{Bb\beta}(\tilde D^{(n)}_k\tilde \psi(x))_{Cc\gamma}.
\end{equation}
Cascade operators are then constructed by projecting to $I=\frac{1}{2}, I_3=\frac{1}{2}$ \cite{adam}.

We use several different patterns of displacements for our three
quark operators, as summarised in Table \ref{tab:opforms}.  Taking linear combinations of
these elemental operators, we project to the irreps of the
octahedral group. The resulting cascade operators for each irreps are  
shown in table 3.
\begin{table}
\caption[captab]{
The various type of operator, and corresponding quark displacements needed in
elemental operators where displacement indices indicate the
direction of the gauge-covariant displacement for each quark;
$i,j,k=\{0,\pm 1,\pm2,\pm3\}$.} \label{tab:opforms}
\begin{center}
\begin{tabular}{|cl|}
\hline
 Operator type &  Displacement indices\\
\raisebox{0mm}{\setlength{\unitlength}{1mm} \thicklines
\begin{picture}(16,10)
\put(8,6.5){\circle{6}} \put(7,6){\circle*{2}}
\put(9,6){\circle*{2}} \put(8,8){\circle*{2}}
\put(0,0){single-site}
\end{picture}}  & \raisebox{3mm}{$i=j=k=0$ }\\
\raisebox{0mm}{\setlength{\unitlength}{1mm} \thicklines
\begin{picture}(23,10)
\put(7,6.2){\circle{5}} \put(7,5){\circle*{2}}
\put(7,7.3){\circle*{2}} \put(14,6){\circle*{2}}
\put(9.5,6){\line(1,0){4}} \put(0,0){singly-displaced}
\end{picture}}  & \raisebox{3mm}{$i=j=0,\ k\neq 0$} \\
\raisebox{0mm}{\setlength{\unitlength}{1mm} \thicklines
\begin{picture}(26,8)
\put(12,5){\circle{3}} \put(12,5){\circle*{2}}
\put(6,5){\circle*{2}} \put(18,5){\circle*{2}}
\put(6,5){\line(1,0){4.2}} \put(18,5){\line(-1,0){4.2}}
\put(-1,0){doubly-displaced-I}
\end{picture}}  & \raisebox{2mm}{$i=0,\ j=-k,\ k\neq 0$} \\
\raisebox{0mm}{\setlength{\unitlength}{1mm} \thicklines
\begin{picture}(20,13)
\put(8,5){\circle{3}} \put(8,5){\circle*{2}}
\put(8,11){\circle*{2}} \put(14,5){\circle*{2}}
\put(14,5){\line(-1,0){4.2}} \put(8,11){\line(0,-1){4.2}}
\put(-5,0){doubly-displaced-L}
\end{picture}}   & \raisebox{4mm}{$i=0,\ \vert j\vert\neq \vert k\vert,
  \ jk\neq 0$}\\
\raisebox{0mm}{\setlength{\unitlength}{1mm} \thicklines
\begin{picture}(20,12)
\put(10,10){\circle{2}} \put(4,10){\circle*{2}}
\put(16,10){\circle*{2}} \put(10,4){\circle*{2}}
\put(4,10){\line(1,0){5}} \put(16,10){\line(-1,0){5}}
\put(10,4){\line(0,1){5}} \put(-5,0){triply-displaced-T}
\end{picture}}   & \raisebox{4mm}{$i=-j,\ \vert j\vert \neq\vert k\vert,
 \ jk\neq 0$} \\
\raisebox{0mm}{\setlength{\unitlength}{1mm} \thicklines
\begin{picture}(20,12)
\put(10,10){\circle{2}} \put(6,6){\circle*{2}}
\put(16,10){\circle*{2}} \put(10,4){\circle*{2}}
\put(6,6){\line(1,1){3.6}} \put(16,10){\line(-1,0){5}}
\put(10,4){\line(0,1){5}} \put(-5,0){triply-displaced-O}
\end{picture}}   & \raisebox{4mm}{$\vert i\vert \neq \vert j\vert \neq
  \vert k\vert,\ ijk\neq 0$}\\
  \hline
\end{tabular}
\end{center}
\end{table}
\vspace*{-0.1in}
\begin{table}[h]
\caption{Type of operators used and their numbers which project into each row of the each irreducible representation 
of the octahedral group $\Lambda=\{G_1,H,G_2\}$ for $\Xi$.}
\begin{center}
\begin{tabular}{|c| c c c |}
\hline 
Operator type & G1 & G2 & H \\
\hline  
Single site           & 4   &  0  & 3  \\
Singly displaced      & 38  &  14 & 52 \\
Doubly displaced-I    & 36  &  12 & 48 \\
Doubly displaced-L    & 96  &  96 & 192 \\
Triply displaced      & 96  &  96 & 192 \\
\hline 
Total                 & 270 & 218 & 487 \\
\hline
 \end{tabular}
 \end{center}
  \end{table}

\vspace*{-0.45in}

\section{Pruning of operators}
Not all the operators of table 3 have a substantial overlap with the low-lying spectrum, and it would be very difficult to use all of
them in a variational calculation.
Indeed, many of them are
noisy which can be seen by examining the diagonal elements of the correlation matrix. We have utilised a similar procedure for
pruning a best set of operator as discussed in Ref.~\cite{adam}, namely :
\begin{itemize}
\item Check the diagonal effective masses for each irrep and for each parity 
 and sorted those in increasing order of the average jackknife error/signal over the first sixteen non-zero time separations to find out best operators for which signal to noise ratios are larger.
\item Using those operators construct normalised correlator matrices in each representation and find their condition numbers.
\item For various subset of matrices find a matrix with minimum condition number.
\end{itemize}
 However, in contrast to Ref. \cite{adam}, we prune each parity channel for each
irreps. separately, i.e., positive parity operators are pruned independently of the
 negative parity operators as their overlap to the physical states
could be different.
\vspace*{-0.1in}
\section{Results}
\vspace*{-0.1in}
We use $24^3\times 64$ anisotropic lattices with the anisotropic
factor $\xi = 3$.
We have generated 2 flavour dynamical Wilson gauge configurations, 
and for fermions we also use Wilson action. 
 We have two set of configurations at temporal
lattice spacings $a_t^{-1} = 5.556$ GeV and 5.310 GeV. 
The scale is
set with the Sommer parameter, and using this scale the pion masses on these lattices are 416
MeV and 572 MeV respectively. Here we will report results for finer lattice (pion mass at 416 MeV) with 860 configurations.

In Fig. 2 we have plotted effective masses for a good pruned operator
for each irrep and for both parity channels.  The left figure is for
the positive parity and the right figure is for the negative parity. As
expected, the lowest state is for the positive parity G1 which can be
associated with the ground state octet (spin-1/2
$\Xi$(1314)). The next lowest state is the positive parity H which can
be of spin 3/2 or spin 5/2. 
The lowest spin contributing to $G_2$ is spin-5/2.  We find this mass
considerably above the ground state in H; whilst a spin-5/2 state
could yield different masses in the two irreps due to
finite-lattice-spacing effects, the considerable difference between
these ground-states masses leads us to conclude that we are observing
a spin-3/2 state in H, and a spin-5/2 state in $G_2$.

We are also able to extract good signals for negative parity states
which are, in general, difficult to obtain. In Fig. 3 (right), we plot
effective masses for the negative parity ground state of 
G1 for a pruned operator along with the commonly used local
negative parity operator. As one can observe that our pruned operator
has much better overlap than the other.  We believe that due to bigger
size of this state one needs an extended operator, and since we have
included those, our pruned operator has bigger overlap to this state.
The spin identification, as argued above, for negative parity states is not possible as 
all of them almost overlap to each other (figure 2, right).
Another interesting point to note is that for G2, the parity of the
ground state is negative as shown in Fig. 3 (left). If one can associate
 the ground state of G2 with a spin 5/2 state, as argued above, then this
result suggests that for spin 5/2 cascade the ground state is for the
negative parity, which is in opposite ordering than the parity ordering
in spin-1/2 and spin-3/2 baryons.

At a lower pion mass one expects to observe signal for multi-hadron
decay states along with resonance states. For example,
 the $N^{*}(1535)$ decays into a nucleon and pion, and similarly the negative parity spin 1/2
cascade will decay to a multi-hadron state of positive parity cascade
and a pion in a S-wave. In Fig. 4 we plotted effective masses of
ground state of irreps $G_1$ (left) and $H$ (right). The horizontal
lines are the sums of the positive parity ground state mass and pion mass,
corresponding to the non-interacting cascade-pion state in S-wave.
Since these lines are overlapping with the
negative parity observed states we argue that the observed states are
either multi-hadron states or a mixture of multi-hadron and resonance
states.
\begin{figure}
\begin{center}
\includegraphics[width=0.497\textwidth,height=0.33\textwidth,clip=true]{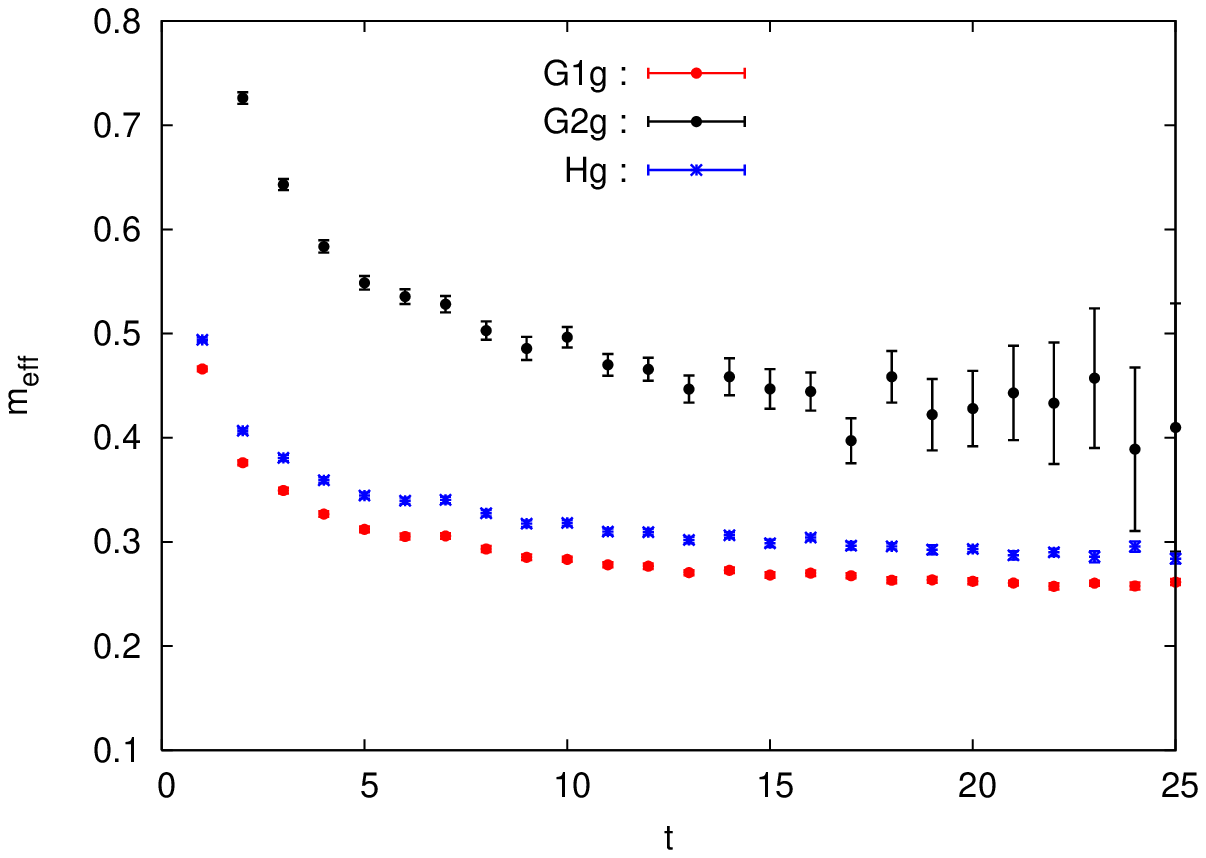}
\includegraphics[width=0.497\textwidth,height=0.33\textwidth,clip=true]{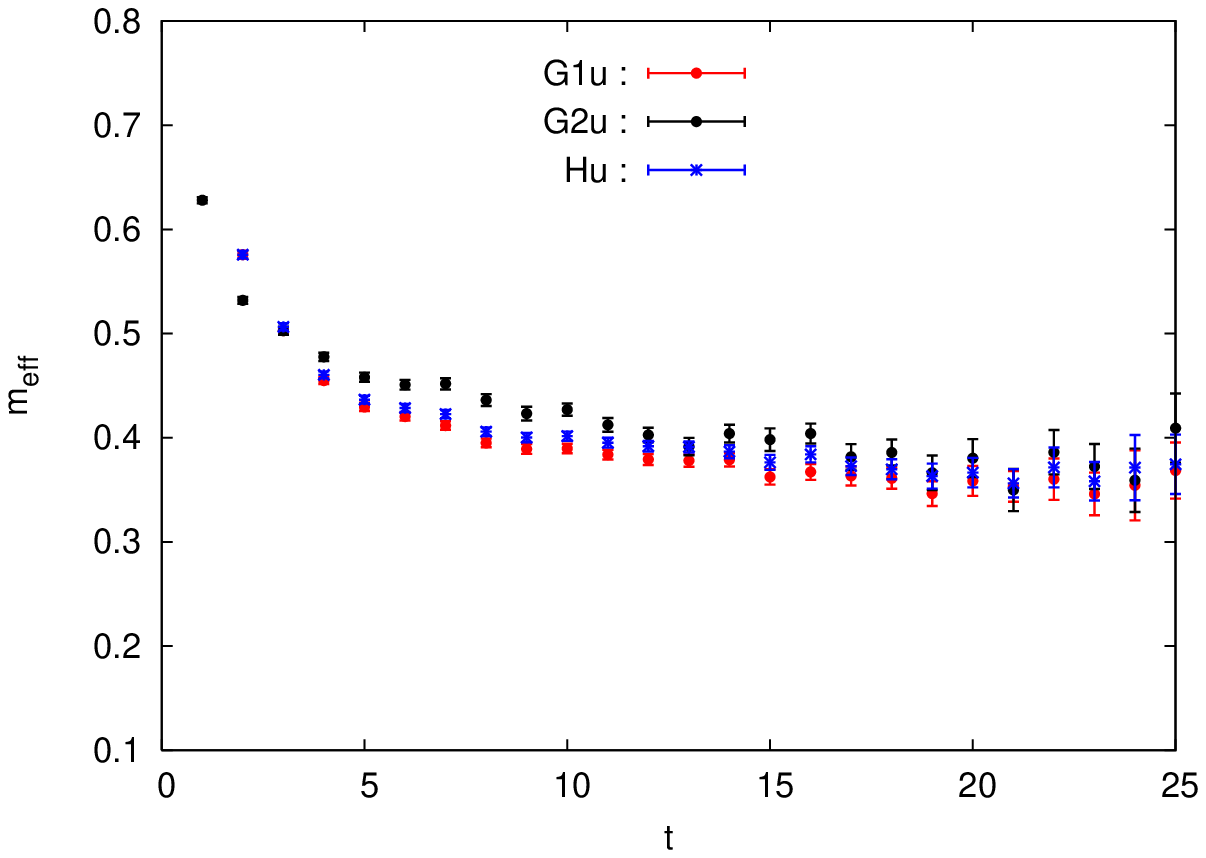} 
\end{center}
\vspace{-0.15in}
\caption{Effective masses of positive (left) and negative (right) parity 
ground state cascade baryons for various irreps for $24^3\times64$ $N_f=2$ lattice QCD data at $m_\pi = 400$ MeV.} \label{fig:eff_all}
\end{figure}
\begin{figure}
\begin{center} 
\includegraphics[width=0.497\textwidth,height=0.33\textwidth,clip=true]{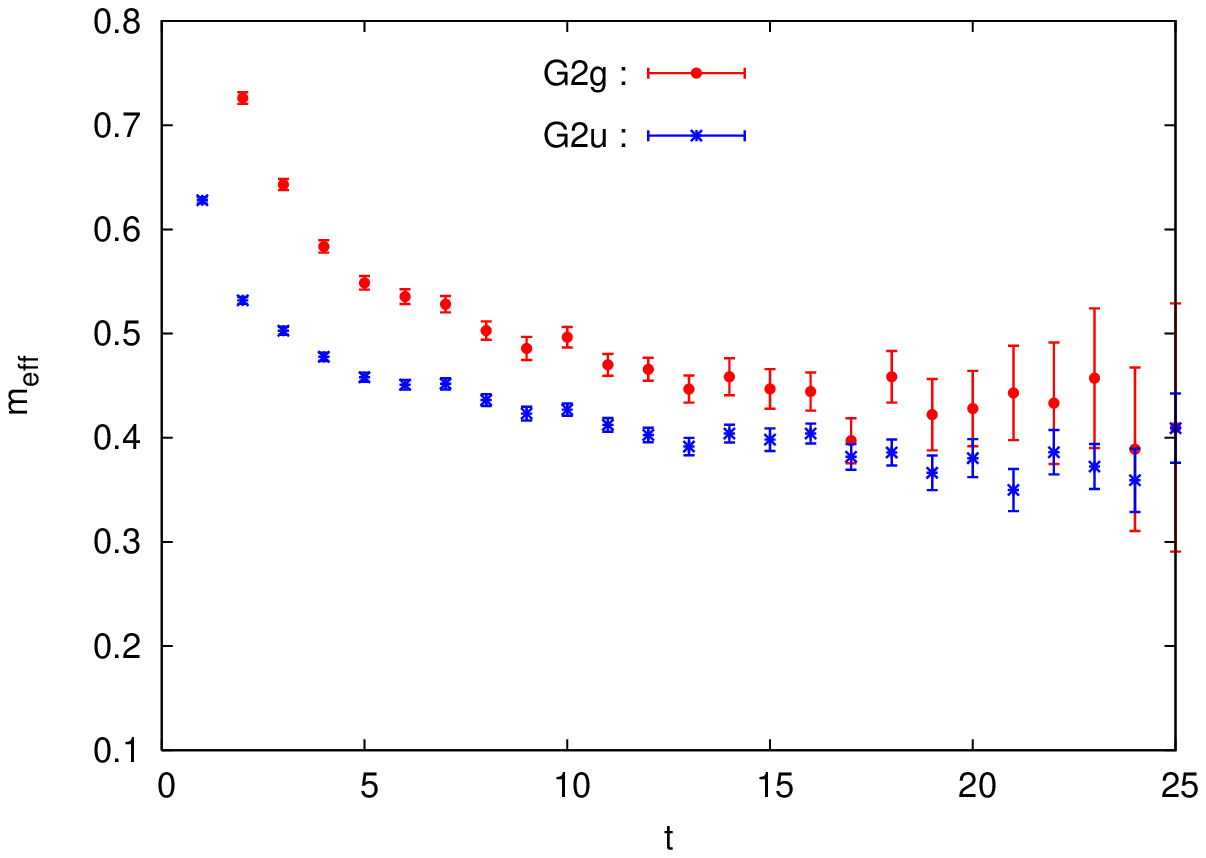}
\includegraphics[width=0.497\textwidth,height=0.33\textwidth,clip=true]{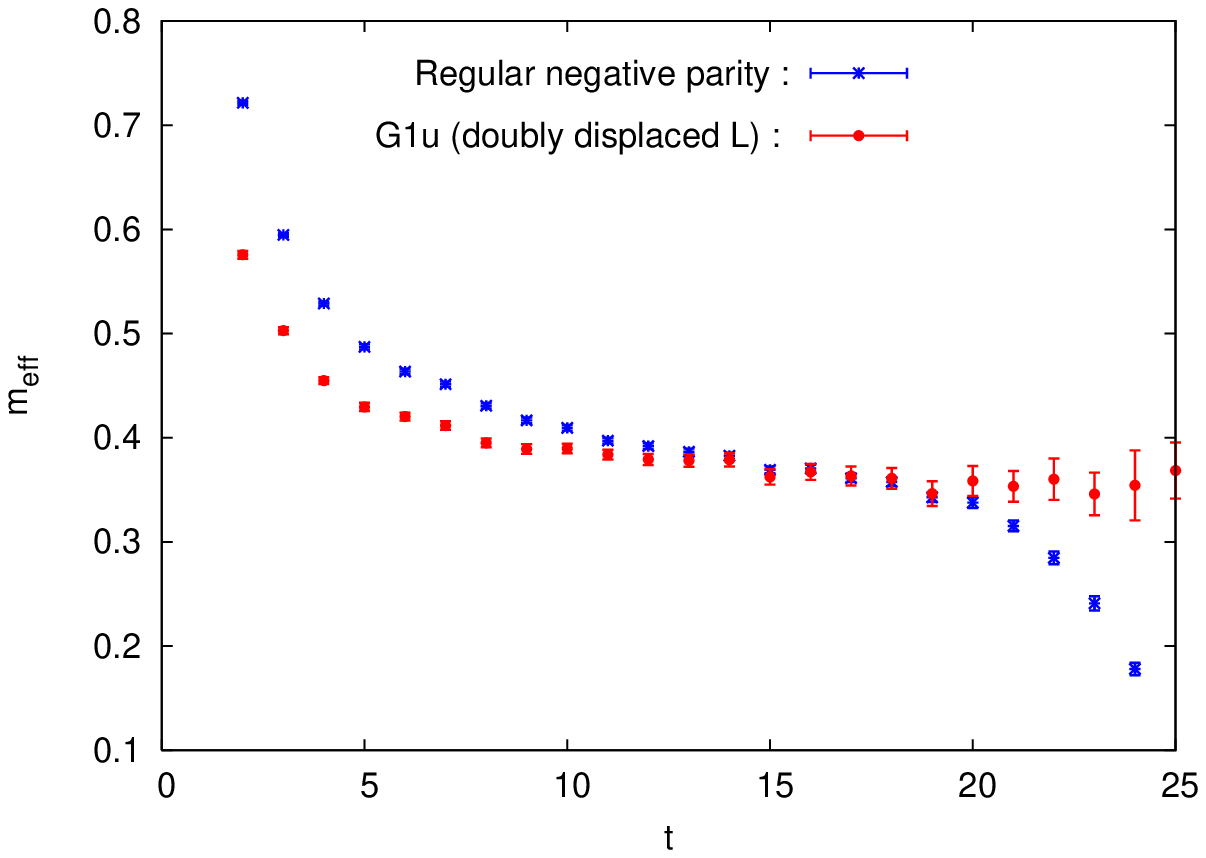}
\end{center}
\vspace{-0.15in}
\caption{Left : A representative effective mass plot for G2 channel showing that the ground state negative parity mass is lower than that of the ground state positive parity. This parity ordering is opposite to what had been seen in other channels. Right : A comparison of effective masses obtained from an extended operator and most used regular local operator for negative parity G1 channel.} \label{fig:eff_G1u}
\end{figure}
\begin{figure}
\begin{center}
\includegraphics[width=0.497\textwidth,height=0.33\textwidth,clip=true]{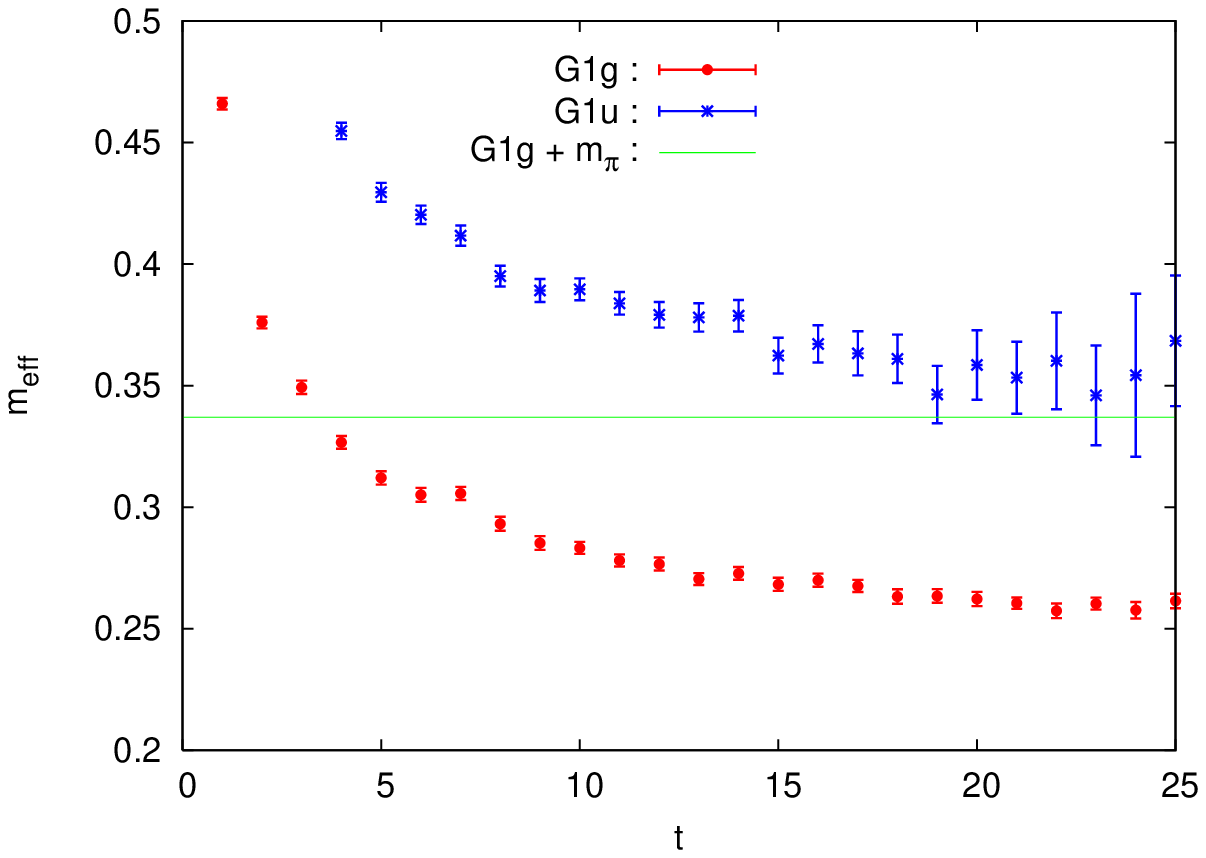}
\includegraphics[width=0.497\textwidth,height=0.33\textwidth,clip=true]{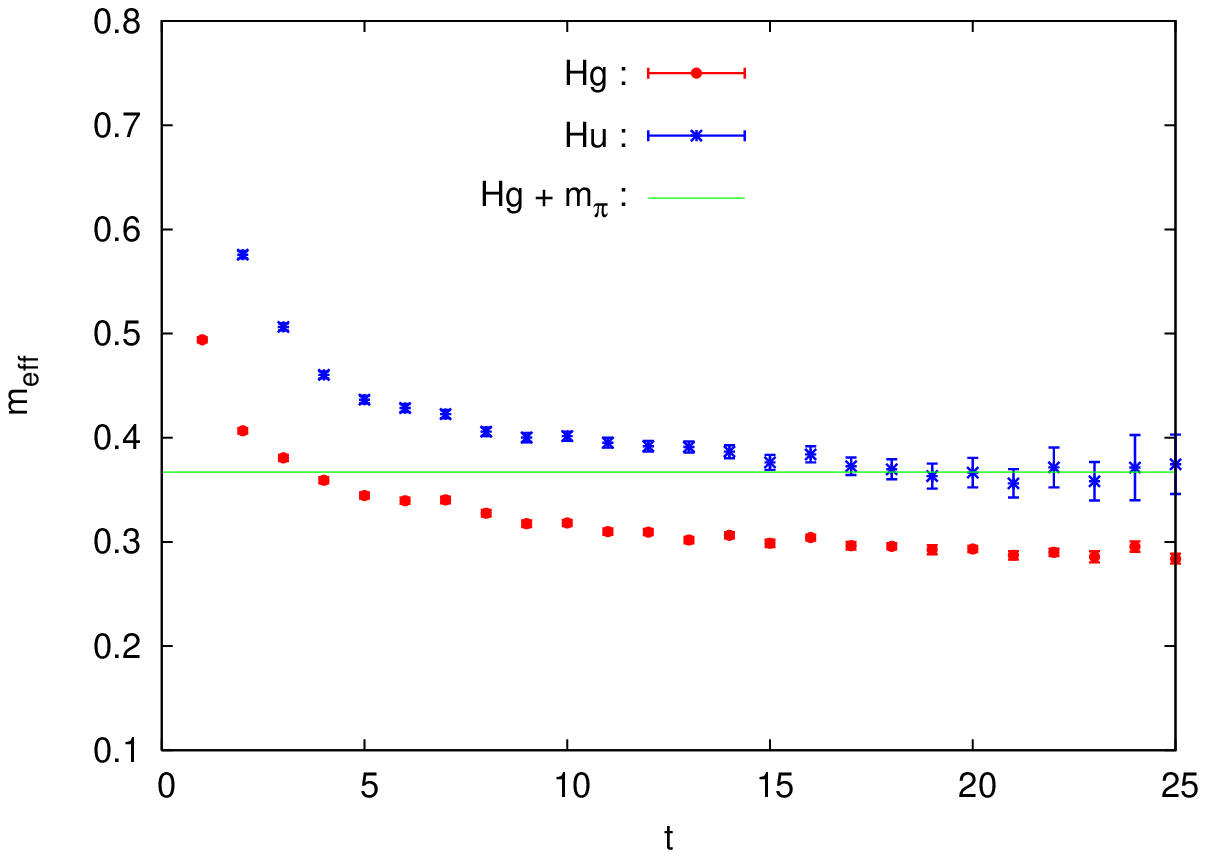} 
\end{center}
\vspace{-0.15in}
\caption{Effective masses of ground state $G_1$ (left) and $H$ (right) irreps. The horizontal lines are sum of the positive parity ground state asses and pion mass which indicate that in these channels the observed negative parity states are probably be the S-wave scattering states.} \label{fig:eff_sct}
\end{figure}

\vspace*{-0.15in}

\section{Conclusions and outlook}
\vspace*{-0.1in}
Given the incomplete experimental picture for cascade baryons, and a comprehensive
lattice study is demanded, we have initiated such a
calculation by using operators which transform as irreps of the
octahedral group. Here we have showed our preliminary results for
$N_f=2$ Wilson fermions at pion mass 416 MeV.  A preliminary analysis
show that we are able to extract clear signals for ground states for
each irrep and for each parity. 
Unlike the case when using only local operators, the use of 
extended operators enables us to extract 
 signals for negative-parity states unambiguously. 
We also believe that, at 416 MeV pion, the negative parity ground
state extracted for irreps G1 and H are probably those of $\Xi-\pi$
multi-hadron scattering states into which the resonance states can decay. 
For the irrep
G2, the negative parity state is lower than that of the positive parity,
which is not common for low lying baryons.  This preliminary study also suggests
that the observed positive parity ground state for irrep H has spin
3/2, whilst that for G2 has spin 5/2.  However, this needs to be
confirmed for other quark masses and then taking continuum limit.  
Our current studies are aimed at carefully delineating the single- and
multi-particle states before we attempt to proceed to extract the
resonance behaviour at lower pion mass We will soon carry out full
variational study to extract excited states as was performed for
ref.~\cite{eric} at this pion mass and at a higher pion mass.  Our
collaboration has begun to generate a series of gauge configurations
for 2+1 dynamical quark flavours for various lattice spacings and
volumes \cite{Nf2+1}.  We will be utilising these lattices for extracting
cascade spectrum along with other hadrons.

\vspace*{-0.1in}

\section{Acknowledgements}
This work was done using the Chroma software
suite~\cite{Edwards:2004sx} on clusters at Jefferson Laboratory using
time awarded under the USQCD Initiative. This research used resources
of the National Center for Computational Sciences at Oak Ridge
National Laboratory, which is supported by the Office of Science of
the U.S. Department of Energy under the US DOE INCITE 2007 Program. This research was supported in
part by the National Science Foundation through Teragrid Resources
provided by the San Diego Supercomputing Center (Blue Gene). 
The research of N.M. is supported under grant No. DST-SR/S2/RJN-19/2007, India.

\end{document}